%% file: main.tex
\documentclass[a4paper,11pt]{article}
\usepackage{pos}
\usepackage{booktabs}
\usepackage{changes}
\usepackage[range-phrase=-, range-units=single]{siunitx}
\usepackage{siunitx}
\DeclareSIUnit\parsec{pc}

\title{First results of low-energy neutrino follow-ups of Run O4 compact binary mergers with the IceCube Neutrino Observatory}

\ShortTitle{O4 neutrino follow-up with IceCube}

\author{The IceCube Collaboration \\{\normalsize \normalfont(a complete list of authors can be found at the end of the proceedings)}\\}

\emailAdd{karlijn.kruiswijk@uclouvain.be}
\emailAdd{mathieu.lamoureux@uclouvain.be}
\emailAdd{gwenhael.dewasseige@uclouvain.be}

\abstract{
We present the results of searches for astrophysical neutrinos of few GeV energy from compact binary mergers detected during the first months of the fourth observing run of the LIGO, Virgo, and KAGRA interferometers. We describe our method, based on a selection of $0.5-5$\,GeV neutrino events in IceCube, where we search for a statistically significant increase in the number of low-energy candidate events detected around the compact binary merger time. With these results, we constrain neutrino-emitting source populations. Finally, we compare our results with constraints set by neutrino searches at $> 10$\,GeV energies and describe the complementarity of these low- and high-energy searches. 

\vspace{4mm}
{\bfseries Corresponding authors:}
Karlijn Kruiswijk$^{*}$, Mathieu Lamoureux, Gwenhaël de Wasseige\\
{\itshape Centre for Cosmology, Particle Physics and Phenomenology - CP3, Universit{\'e} Catholique de Louvain, Louvain-la-Neuve, Belgium}\\[4mm]
$^*$ Presenter

\ConferenceLogo{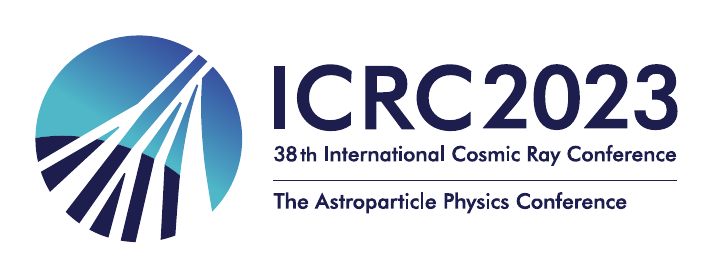}

\FullConference{The 38th International Cosmic Ray Conference (ICRC2023)\\ 26 July -- 3 August, 2023\\ Nagoya, Japan}
}

\begin{document}

\maketitle

\section{Introduction}
\label{sec1}

In the past ten years, the detections of astrophysical neutrinos by IceCube and gravitational wave sources with the LIGO and Virgo interferometers have opened a new window to the Universe. Unlike photons, these messengers are not stopped by dense environments or by dust, hence providing complementary information on source mechanisms and underlying processes.

In the years since the first detection of a binary black hole merger~\cite{LIGOScientific:2016aoc}, many more GW detections have followed, including neutron star - black hole and binary neutron star mergers. The ongoing fourth observing run (O4) has started with a horizon of $\qtyrange{130}{150}{\mega\parsec}$ for binary neutron star mergers, though it still aims to reach $\qtyrange{160}{190}{\mega\parsec}$ with improvements in the detector hardware as the run proceeds~\cite{O4}.

For every GW alert from LIGO and Virgo, the astroparticle community has carried out follow-up observations~\cite{abbasi_2021GeVmergers}. An excess of neutrinos has not yet been observed, but such non-observations provide constraints on the multi-messenger picture of these merger events and on the involved processes, such as the acceleration mechanism of hadronic particles and the composition of the merged objects. 

The prime example of multi-messenger astronomy is the detection of a gamma-ray burst (GRB) from the binary neutron star merger GW170817~\cite{Abbott_2017,LIGOScientific:2017ync}. With the discovery of this GW+GRB coincidence and as neutrinos have long been expected from GRBs, there has been an increased interest in neutrino production in merger events. While these neutrinos are typically expected to have energies in the TeV--PeV range~\cite{Waxman1997} or in the $\qtyrange{10}{100}{\giga\electronvolt}$ range~\cite{Murase_2013}, GeV neutrinos can also be produced by proton-proton or proton-neutron interactions in the denser medium of the burst~\cite{Murase_2013}. Therefore, observation of such neutrinos would not only highlight the presence of hadron acceleration but also give a deeper look into the merger events.

The time of neutrino emission from merger events is not tightly constrained. Neutrinos can originate from the GRB following the merger, and in this case there is an uncertainty in the time of emission. Both the timing of the GRB after the merger event and when the neutrino is emitted from the GRB is unknown. There is even the possibility of neutrinos being emitted during a precursor, where the neutrino emission can happen before the GRB. Therefore, searches often consider a conservative time window of $\pm \qty{500}{\second}$ around the merger time~\cite{BARET20111}.

The IceCube Neutrino Observatory, a Cherenkov detector located at the South Pole, can be used to search for these neutrinos. It consists of an array of more than 5000 PMTs distributed along vertical strings instrumenting a cubic kilometer of ice. Neutrino interactions are identified when their secondary charged particles induce Cherenkov light in the ice. The detector geometry is optimized for detecting TeV--PeV neutrinos. 
The detection of lower-energy events is realized by using the DeepCore sub-array located at the center of IceCube. This sub-array consists of more densely placed PMTs that are more sensitive than the standard IceCube PMTs. With DeepCore, it is possible to search for astrophysical neutrinos between $\qtyrange{0.5}{5}{\giga\electronvolt}$ with a specialized selection procedure called ELOWEN~\cite{TalkELOWEN}.

The ELOWEN selection can reduce the initial rate of triggered events from the \unit{\kilo\hertz} level~\cite{Aartsen_2017} down to \qty{0.02}{\hertz}, as detailed in~\cite{Abbasi_2021GeVSolar}. Before the ELOWEN selection, the data is dominated by atmospheric muons and "noise events", including uncorrelated thermal noise, uncorrelated radioactive noise, and correlated scintillation noise~\cite{Abbasi_2010,ABBASI2009294}. Using several different filtering steps, ELOWEN can reduce these noise sources dramatically while ensuring about 40\% selection efficiency for GeV neutrinos~\cite{Abbasi_2021GeVSolar,TalkELOWEN}. This final sample is still dominated by noise events with a subdominant contribution from atmospheric neutrinos, which are estimated from simulations to occur at the \unit{\milli\hertz} level.

Despite the large background and lack of direction reconstruction, searching for neutrinos from transient events with ELOWEN is still possible by identifying an excess in the number of events in a short period around the transient detection. The number of neutrino events can be compared to the expected background to extract the observation significance and eventually put constraints on the flux. With this method, analyses using ELOWEN have been targeting neutrinos from solar flares~\cite{Abbasi_2021GeVSolar}, as well as from GRB 221009A, the brightest GRB ever observed~\cite{GRB22paper,TalkGRB22}.

\section{Analysis method}
\label{sec2}

In these proceedings, ELOWEN is used to search for neutrinos originating from GW sources. A time window of \qty{1000}{\second} centered on the GW time is employed as prescribed in~\cite{BARET20111} for such studies. A special treatment is made for mergers that are reported with $> 50\%$ probability to involve at least one neutron star, as they are more likely to be associated with a GRB. Based on the first detected binary neutron star (BNS) merger, GW170817, and its associated GRB detected \qty{1.7}{\second} later by Fermi-GBM and Swift~\cite{Abbott_2017}, we define $[t_{\rm GW}, t_{\rm GW} + \qty{3}{\second}]$ as an alternative time window for the search of a prompt signal, where $t_{\rm GW}$ is the merger time. 

Several checks are performed to ensure neutrino data quality during the gravitational wave event. In addition to the automated data quality checks performed in IceCube~\cite{Aartsen_2017} and prior to the unblinding of the data, an \qty{8}{\hour} time window directly before the chosen time window is analyzed to estimate the ELOWEN background rate and ensure it is compatible with the expected \qty{20}{\milli\hertz} (typically falling within the $\qtyrange{17}{23}{\milli\hertz}$ range given the statistical uncertainty on the estimation). After unblinding, the possibility of localized noise events in a single PMT or string (due to uncaught detector anomalies) is excluded by checking the spatial distribution of the neutrino candidate events within IceCube during the \qty{1000}{\second} time window.

If all checks are passed, the observed number of events in the search time window $N_{\rm on}$ is compared to the background estimated from the \qty{8}{\hour}. 
The Li \& Ma approach~\cite{Li:1983fv} is used to quantify the significance of the observation (in units of $\sigma$) and the related 90\% sensitivity is obtained by computing which signal strength is needed to achieve a significance corresponding to a $\sim 1.64\sigma$ excess.
In the absence of any significant excess ($>3\sigma$), a Bayesian method is employed to compute the individual 90\% upper limits on the number of signal events and the corresponding all-flavor time-integrated flux~\cite{Knoetig:2014dha}. The following likelihood is defined:
\begin{equation}
    \mathcal{L}(N_{\rm on}, N_{\rm off} \vert N_{\rm sig}, N_{\rm bkg}, \alpha) = \textrm{Poisson}(N_{\rm on}; N_{\rm sig} + N_{\rm bkg}) \times \textrm{Poisson}(N_{\rm off}; N_{\rm bkg}/\alpha),
\end{equation}
where $N_{\rm on}$ ($N_{\rm off}$) is the observed number of events in the search (\qty{8}{\hour}) time window, $\alpha$ is the ratio between the livetimes in the search and \qty{8}{\hour} time windows, and $N_{\rm sig}$ ($N_{\rm bkg}$) is the estimated number of signal (background) events in the search time window. A Jeffreys uninformative prior~\cite{Jeffreys:1945} is then used to get the corresponding posterior distribution. The latter is then marginalized over $N_{\rm bkg}$ to finally compute the 90\% upper limit $N^{90\%}_{\rm sig}$.

\begin{figure}[hbtp]
    \centering
    \includegraphics[width=0.9\linewidth]{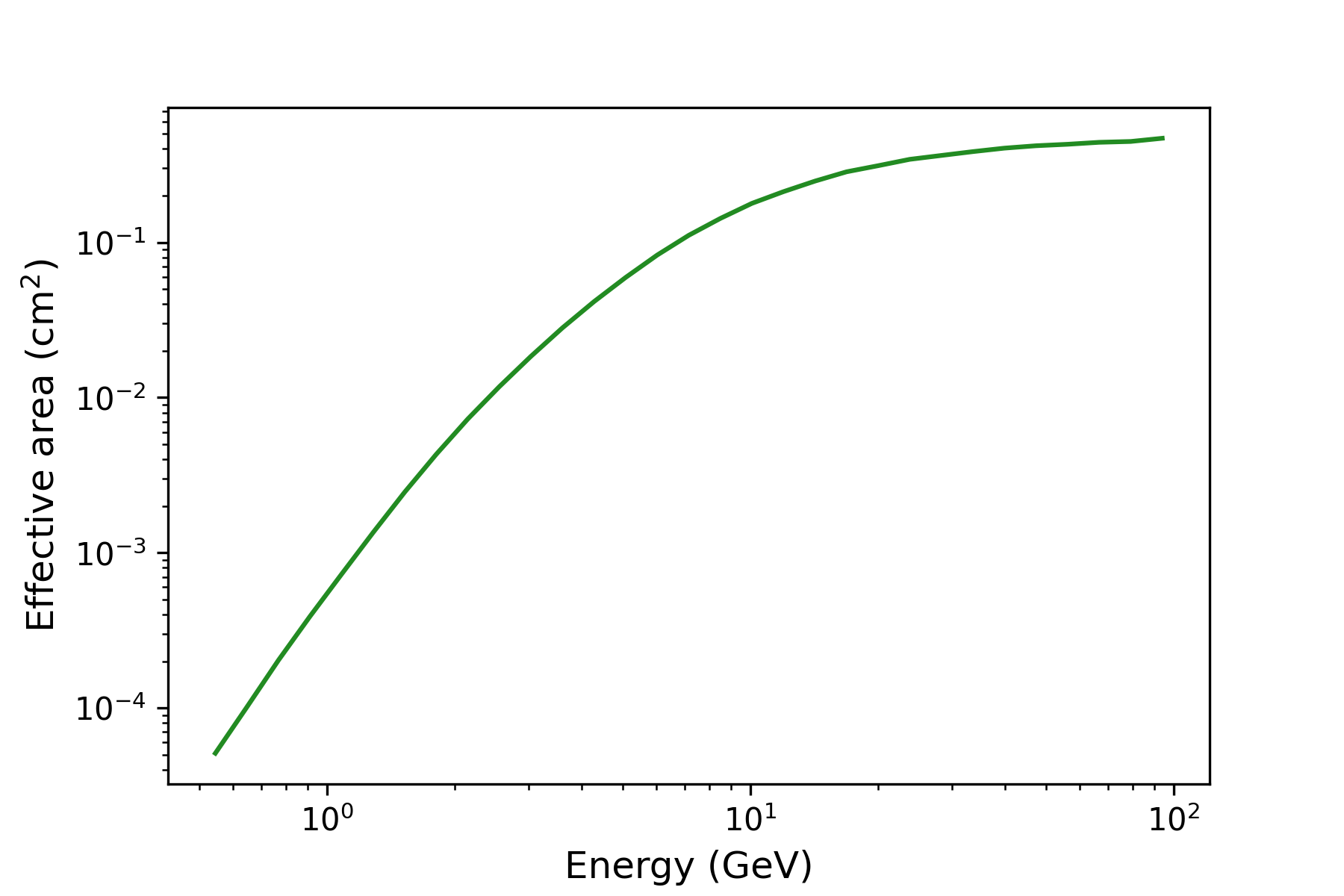}
    \caption{All-flavor effective area for the ELOWEN selection from $0.5$ to $\qty{100}{\giga\electronvolt}$.}
    \label{figEff_area}
\end{figure}

The conversion to a flux limit uses the summed effective area for all neutrino flavors $A_{\rm eff}(E)$ ranging from \num{0.5} to \qty{100}{\giga\electronvolt}, as shown in \autoref{figEff_area}. Only power-law spectra are considered:
\begin{equation}
    \dfrac{{\rm d}N}{{\rm d}E} = \phi \left(\dfrac{E}{\unit{\giga\electronvolt}} \right)^{-\gamma},
\end{equation}
where $\phi$ is the flux normalization in \unit{\per\giga\electronvolt\per\square\centi\meter} and $\gamma = 2.0, 2.5, 3.0$. Limits are then reported in terms of  $\phi$: 
\begin{equation}
    \phi^{90\%} = \dfrac{6 \times N^{90\%}_{\rm sig}}{\int_{\qty{0.5}{\giga\electronvolt}}^{\qty{100}{\giga\electronvolt}} A_{\rm eff}(E) \times (E/\unit{\giga\electronvolt})^{-\gamma} {\rm d}E},
\end{equation}
where the factor $6$ is used to convert from single-flavor to all-flavor flux. If ones replaces the units with \unit{\giga\electronvolt\per\square\centi\meter}, it is the same as the limit on $E^2 {\rm d}N/{\rm d}E$ at \qty{1}{\giga\electronvolt}.

\section{First results}
\label{sec3}

The follow-up analysis has been performed for nine GW significant alerts during the engineering run ER15 and the beginning of O4, including two neutron star - black hole (NSBH) mergers while the rest are most likely binary black hole mergers, as reported in the alert GCN notices.

\begin{figure}[hbtp]
    \centering
    \includegraphics[width=0.8\linewidth]{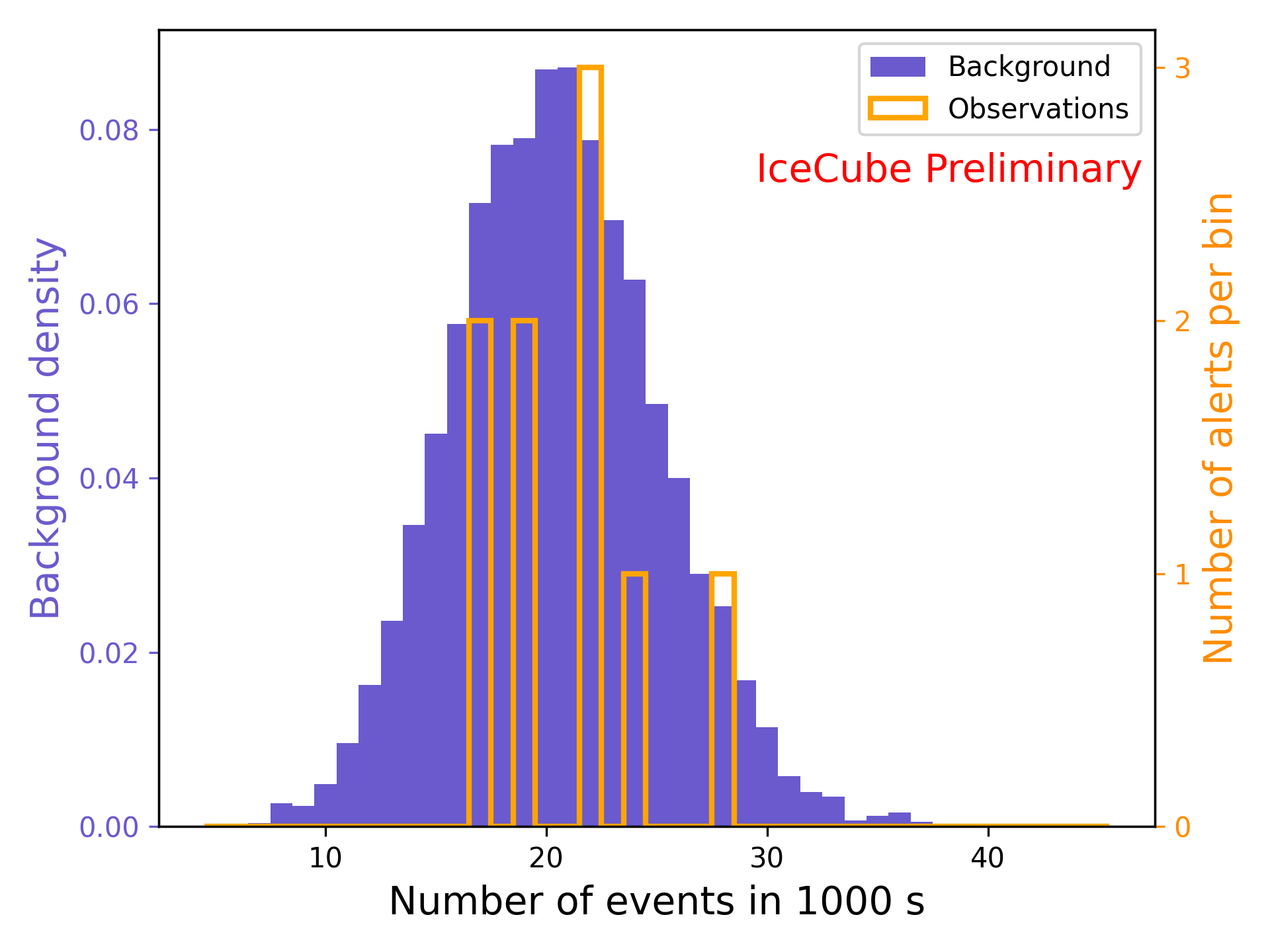}
    \caption{Distribution of the observed number of events in the \qty{1000}{\second} time window for the O4 GW alerts (orange histogram) compared with the background expectation (blue filled histogram) as estimated from the equivalent of one month of data taken during periods when no transient sources (GWs, GRBs, high-energy neutrinos, solar flares) were reported on GCN or ATel.}
    \label{fig:results}
\end{figure}

The results in terms of the number of events are summarized in \autoref{fig:results}. Given the average expected background rate of $\sim \qty{20}{\milli\hertz}$, the 90\% flux sensitivity of the search computed with the Li \& Ma method is $\phi < \qty{1.1e3}{\per\giga\electronvolt\per\square\centi\meter}$ for an $E^{-2}$ spectrum and the \qty{1000}{\second} time window. No event has a significance higher than $3\sigma$ and individual upper limits are then computed using the Bayesian method described in the previous section. They are reported in \autoref{tab:results} for the \qty{1000}{\second} time window and in \autoref{tab:results3s} for the \qty{3}{\second} time window for the two NSBH candidates.

\begin{table}[hbtp]
    \centering
    \begin{tabular}{cl|ccS[table-format=2.1]|*{3}{S[table-format=1.1e2,table-auto-round,table-number-alignment=center,retain-zero-exponent]}} 
        \toprule
         & & & & & \multicolumn{3}{c}{Upper limits on $\phi$ [\unit{\per\giga\electronvolt\per\square\centi\meter}]} \\ 
         & & {$\langle N_{\rm bkg} \rangle$} & {$N_{\rm on}$} & {$N^{90\%}_{\rm sig}$} & {$\gamma = 2$} & {$\gamma = 2.5$} & {$\gamma = 3$} \\
        \midrule
        S230518h & [\href{https://gcn.gsfc.nasa.gov/notices_l/S230518h.lvc}{GCN}] & 18.81 & 24 & 12.7 & 1.70e+03 & 5.45e+03 & 1.38e+04 \\ 
        S230520ae & [\href{https://gcn.gsfc.nasa.gov/notices_l/S230520ae.lvc}{GCN}] & 19.06 & 22 & 10.6 & 1.42e+03 & 4.56e+03 & 1.16e+04 \\ 
        S230522a & [\href{https://gcn.gsfc.nasa.gov/notices_l/S230522a.lvc}{GCN}] & 19.36 & 19 & 8.1 & 1.08e+03 & 3.48e+03 & 8.83e+03 \\ 
        S230522n & [\href{https://gcn.gsfc.nasa.gov/notices_l/S230522n.lvc}{GCN}] & 19.82 & 17 & 6.7 & 8.92e+02 & 2.86e+03 & 7.27e+03 \\ 
        S230529ay & [\href{https://gcn.gsfc.nasa.gov/notices_l/S230529ay.lvc}{GCN}] & 20.34 & 19 & 7.6 & 1.02e+03 & 3.27e+03 & 8.32e+03 \\ 
        S230601bf & [\href{https://gcn.gsfc.nasa.gov/notices_l/S230601bf.lvc}{GCN}] & 17.67 & 28 & 17.9 & 2.40e+03 & 7.69e+03 & 1.96e+04 \\ 
        S230605o & [\href{https://gcn.gsfc.nasa.gov/notices_l/S230605o.lvc}{GCN}] & 19.74 & 22 & 10.2 & 1.36e+03 & 4.36e+03 & 1.11e+04 \\ 
        S230606d & [\href{https://gcn.gsfc.nasa.gov/notices_l/S230606d.lvc}{GCN}] & 19.06 & 17 & 7.0 & 9.31e+02 & 2.98e+03 & 7.59e+03 \\ 
        S230609u & [\href{https://gcn.gsfc.nasa.gov/notices_l/S230609u.lvc}{GCN}] & 19.29 & 22 & 10.5 & 1.41e+03 & 4.52e+03 & 1.15e+04 \\ 
        \bottomrule
    \end{tabular}
    \caption{Summary of follow-up results for the first O4 GW alerts. The first column indicates the alert name and the link to the corresponding GCN notices. The second and third columns report the numbers of events in the \qty{1000}{\second} time window expected from background $\langle N_{\rm bkg} \rangle = \alpha N_{\rm off}$ and observed $N_{\rm on}$. The fourth column contains the 90\% upper limit on the number of signal events, and the last three columns are the corresponding 90\% upper limits on the all-flavor flux normalization $\phi$ for different spectral indices.}
    \label{tab:results}
\end{table}

\begin{table}[hbtp]
    \centering
    \begin{tabular}{cl|ccS[table-format=2.1]|*{3}{S[table-format=1.1e2,table-auto-round,table-number-alignment=center,retain-zero-exponent]}} 
        \toprule
         & & & & & \multicolumn{3}{c}{Upper limits on $\phi$ [\unit{\per\giga\electronvolt\per\square\centi\meter}]} \\ 
         & & {$\langle N_{\rm bkg} \rangle$} & {$N_{\rm on}$} & {$N^{90\%}_{\rm sig}$} & {$\gamma = 2$} & {$\gamma = 2.5$} & {$\gamma = 3$} \\
        \midrule
        S230518h & [\href{https://gcn.gsfc.nasa.gov/notices_l/S230518h.lvc}{GCN}] & 0.04 & 0 & 1.5 & 2.04e+02 & 6.54e+02 & 1.66e+03 \\ 
        S230529ay & [\href{https://gcn.gsfc.nasa.gov/notices_l/S230529ay.lvc}{GCN}] & 0.04 & 0 & 1.5 & 2.05e+02 & 6.57e+02 & 1.67e+03 \\ 
        \bottomrule
    \end{tabular}
    \caption{Summary of follow-up results for the first O4 GW alerts with $>50\%$ probability of involving a neutron star. The first column indicates the alert name and the link to the corresponding GCN notices. The second and third columns report the numbers of events in the \qty{3}{\second} time window expected from background $\langle N_{\rm bkg} \rangle = \alpha N_{\rm off}$ and observed $N_{\rm on}$. The fourth column contains the 90\% upper limit on the number of signal events, and the last three columns are the corresponding 90\% upper limits on the all-flavor flux normalization $\phi$ for different spectral indices.}
    \label{tab:results3s}
\end{table}

\section{Summary and Perspectives}

With the start of O4, we have started presenting our observations of its merger candidates and their upper limits on the neutrino flux at GeV energies using the ELOWEN selection at the IceCube Neutrino Telescope. Because no significant excess was found, neither in a \qty{1000}{\second} nor in a \qty{3}{\second} time window, we have set upper limits on the neutrino emission at GeV scale from these GW sources, complementing higher-energy neutrino searches~\cite{Thwaites:2023icrc}. For an $E^{-2}$ spectrum, the limits are $3-4$ orders of magnitude higher than the ones at higher energies. However, the ELOWEN search can probe softer spectra ($E^{-2.5}$, $E^{-3}$) for which more GeV-scale neutrinos are expected.

Moreover, we are currently working to improve the sensitivity of the ELOWEN selection~\cite{TalkELOWEN}. The planned IceCube Upgrade will add seven strings to the center of the detector~\cite{Ishihara:2019aao}. Due to the dense PMT spacing, this will also improve the sensitivity in the \unit{\giga\electronvolt} region.

The observations will continue for the many GW events expected during O4 and future runs. Combined with the improvements on the selection and detector sides, this may allow us to carry out detailed source population studies. For instance, we may constrain the typical neutrino emission from sub-populations of similar objects (e.g. Binary Black Hole mergers with relatively large spins), and therefore better understand the immediate environment around such sources.

\bibliographystyle{ICRC}
\bibliography{references}

\clearpage

\input{authorlist_IceCube.tex}

\end{document}

%% file: authorlist_IceCube.tex
\section*{Full Author List: IceCube Collaboration}

\scriptsize
\noindent
R. Abbasi$^{17}$,
M. Ackermann$^{63}$,
J. Adams$^{18}$,
S. K. Agarwalla$^{40,\: 64}$,
J. A. Aguilar$^{12}$,
M. Ahlers$^{22}$,
J.M. Alameddine$^{23}$,
N. M. Amin$^{44}$,
K. Andeen$^{42}$,
G. Anton$^{26}$
% ...
\\
\\
% ...
$^{12}$ Universit{\'e} Libre de Bruxelles, Science Faculty CP230, B-1050 Brussels, Belgium \\
% ...
$^{17}$ Department of Physics, Loyola University Chicago, Chicago, IL 60660, USA \\
$^{18}$ Dept. of Physics and Astronomy, University of Canterbury, Private Bag 4800, Christchurch, New Zealand \\
% ...
$^{22}$ Niels Bohr Institute, University of Copenhagen, DK-2100 Copenhagen, Denmark \\
$^{23}$ Dept. of Physics, TU Dortmund University, D-44221 Dortmund, Germany \\
% ...
$^{26}$ Erlangen Centre for Astroparticle Physics, Friedrich-Alexander-Universit{\"a}t Erlangen-N{\"u}rnberg, D-91058 Erlangen, Germany \\
% ...
$^{40}$ Dept. of Physics and Wisconsin IceCube Particle Astrophysics Center, University of Wisconsin{\textendash}Madison, Madison, WI 53706, USA \\
% ...
$^{42}$ Department of Physics, Marquette University, Milwaukee, WI, 53201, USA \\
% ...
$^{44}$ Bartol Research Institute and Dept. of Physics and Astronomy, University of Delaware, Newark, DE 19716, USA \\
% ...
$^{63}$ Deutsches Elektronen-Synchrotron DESY, Platanenallee 6, 15738 Zeuthen, Germany  \\
$^{64}$ Institute of Physics, Sachivalaya Marg, Sainik School Post, Bhubaneswar 751005, India \\
% ...

\subsection*{Acknowledgements}

\noindent
The authors gratefully acknowledge the support from the following agencies and institutions:
USA {\textendash} U.S. National Science Foundation-Office of Polar Programs,
U.S. National Science Foundation-Physics Division,
U.S. National Science Foundation-EPSCoR,
Wisconsin Alumni Research Foundation,
Center for High Throughput Computing (CHTC) at the University of Wisconsin{\textendash}Madison,
Open Science Grid (OSG),
Advanced Cyberinfrastructure Coordination Ecosystem: Services {\&} Support (ACCESS),
Frontera computing project at the Texas Advanced Computing Center,
U.S. Department of Energy-National Energy Research Scientific Computing Center,
Particle astrophysics research computing center at the University of Maryland,
Institute for Cyber-Enabled Research at Michigan State University,
and Astroparticle physics computational facility at Marquette University;
Belgium {\textendash} Funds for Scientific Research (FRS-FNRS and FWO),
FWO Odysseus and Big Science programmes,
and Belgian Federal Science Policy Office (Belspo);
Germany {\textendash} Bundesministerium f{\"u}r Bildung und Forschung (BMBF),
Deutsche Forschungsgemeinschaft (DFG),
Helmholtz Alliance for Astroparticle Physics (HAP),
Initiative and Networking Fund of the Helmholtz Association,
Deutsches Elektronen Synchrotron (DESY),
and High Performance Computing cluster of the RWTH Aachen;
Sweden {\textendash} Swedish Research Council,
Swedish Polar Research Secretariat,
Swedish National Infrastructure for Computing (SNIC),
and Knut and Alice Wallenberg Foundation;
European Union {\textendash} EGI Advanced Computing for research;
Australia {\textendash} Australian Research Council;
Canada {\textendash} Natural Sciences and Engineering Research Council of Canada,
Calcul Qu{\'e}bec, Compute Ontario, Canada Foundation for Innovation, WestGrid, and Compute Canada;
Denmark {\textendash} Villum Fonden, Carlsberg Foundation, and European Commission;
New Zealand {\textendash} Marsden Fund;
Japan {\textendash} Japan Society for Promotion of Science (JSPS)
and Institute for Global Prominent Research (IGPR) of Chiba University;
Korea {\textendash} National Research Foundation of Korea (NRF);
Switzerland {\textendash} Swiss National Science Foundation (SNSF);
United Kingdom {\textendash} Department of Physics, University of Oxford.